# Free energy of a two-liquid system of charge carriers in strongly coupled electron and phonon fields and common nature of three phases in hole-doped cuprates

A.E. Myasnikova[*], S. V. Doronkina, R. R. Arutyunyan and A. H. Dzhantemirov

Southern federal university, 5 Zorge str., 344090 Rostov-on-Don, Russia

[*]Corresponding author, e-mail myasnikova67@yandex.ru

Hole-doped cuprates demonstrate partially coexisting pseudogap, charge density wave (CDW) and superconducting phases. Recently their common nature was supposed due to similar doping dependence of their critical temperatures. In the CDW phase, a large frozen deformation of strongly ionic lattice is observed. Like for a single carrier in a highly polarizable lattice, for multiple carriers in it the choice between delocalized states of all the carriers and autolocalized states of some part of them is determined by minimum of free energy of interacting carrier and phonon fields. Applying variational approach, we calculate free energy of a two-liquid system of carriers with cuprates-like dispersion comprising liquid of autolocalized carriers (large polarons and bipolarons) and Fermi liquid of delocalized carriers. Comparing it with the free energy of pure Fermi liquid and calculating (with standard methods of Bose-liquid theory) a temperature of superfluid transition in the large-bipolaron liquid we identify areas with presence of pseudogap (caused by impact of (bi)polaron potential on delocalized quasiparticles), CDW and superconductivity in a phase diagram. They are in the same places as in hole-doped cuprates, and similarly to cuprates the shape of the calculated phase diagram is stable with respect to wide-range change of the system characteristics. Like in cuprates, the calculated temperature of the superconducting transition increases with the number of conducting planes in the unit cell, the calculated superfluid density decreases with doping at overdoping, and the bipolaron density (together with bipolaronic plasmon energy) saturates at optimal doping. The results obtained allow us to discuss ways of increasing the temperature of the superfluid transition in large-bipolaron liquid and open a possibility of studying the current-carrying state and properties of bipolaron condensate.

## I. INTRODUCTION

Rich phase diagram of hole-doped cuprates comprising partially coexisting pseudogap (PG), charge ordering (CO), or charge density wave (CDW), and superconducting (SC) phases amazes imagination. Even more exciting is invariable presence of these phases in various cuprate superconductors, that hints to their common nature. Relation between the CDW and Fermi surface (FS) reconstruction in cuprates resulting in the PG formation was initially supposed following conventional CDW model. However, experimental studies of the PG manifestation in ARPES spectra of cuprates demonstrated a distinct from the conventional CDW picture known as particle-hole asymmetry [4,5]. Interestingly, that tunneling investigation of CO also demonstrates electron-hole asymmetry: the CO was observed only at tunneling electrons into the sample from the STM tip and not at opposite tunneling [6], in a suggested approach these asymmetries arise naturally.

Recently common origin of the PG, CDW and SC phases was supposed on the base of similar doping behavior of their critical temperatures [7]. Supporting this assumption, the CO peak was observed in resonant X-ray scattering (RXS) spectra of overdoped cuprates with p=0.21 (disappearing at p=0.25) [8], and concurrent disappearance of the SC and CO at close doping was reported [9]. There are also three other coincidences of experimentally observed characteristics universal for cuprate superconductors that are not accident in the suggested approach. The first of them is close values of the CO wavelength and the SC coherence length, or the pair size. Relation of the CDW and PG with a pair formation was assumed in the pair density wave (PDW) model [10,11], where the CDW and PG emerge due to pairing of delocalized carriers with non-opposite momentums. The PDW model reproduces the particle-hole asymmetry of the antinodal PG. However, it suggests two different pairings - Amperean antinodal and d-wave nodal - supposing (without calculation) that free energy of the system in these phases is close. The question arising is why the antinodal pairing with high characteristic energy of the order of PG width does not result in superconductivity with very high transition temperature.

Here we calculate the free energy of a somewhat other two-liquid system of carriers in which the pair size is also related with the CDW period and compare it with the free energy of pure Fermi-liquid of charge carriers. The considered two-liquid system comprises a liquid of large polarons and bipolarons (for brevity it is called below a (bi)polaron liquid) coexisting with the Fermi liquid of delocalized carriers. In distinct from the PDW model in such approach there is only one pairing and PG is not a pairing gap but emerges due to impact of the potential created by (bi)polarons on stationary states of delocalized carriers.

The reason to study such two-component system of carriers is large frozen lattice deformation in the CDW phase of cuprates observed directly with X-ray diffraction [12] and indirectly through giant softening of the phonon modes with the wave vectors close to the CO one [13,14]. Let us stress that observed lattice deformation in the cuprates' CDW phase is not transient as in a Cooper pair but is classical one (with non-zero average values), like in a polaron in which it is especially evident in the coherent-states representation [15]. Both in a CDW and in a polaron forming in strongly polarizable lattice the lattice deformation is concurrently formed by carriers and represents potential wells for them, the well size is much larger than the lattice constant, so that both these problems are studied using Fröhlich Hamiltonian of the electron-phonon interaction (EPI) [16-18]. As was noted by Landau [19], if the polarization potential well is sufficiently deep for a discrete level of a carrier to occur in it then in the ground state the carrier is autolocalized. The well depth in the ground state is determined by minimizing the energy of interacting electron and phonon fields over the parameters of the phonon vacuum deformation and of the carrier wave function (provided the adiabatic condition is satisfied) [20,15].

Obviously, this conclusion can be generalized to a system of multiple carriers and polarization potential wells. Broad bands in ARPES and optical spectra and absence of Fermi crossing in ARPES spectra at doping $p<0.1$ [21-29] indicate presence of autolocalized carriers in cuprates at low temperatures. Coexistence of localized (small (bi)polarons) and delocalized carriers in cuprates was supposed earlier on the base of doping behavior of susceptibility [30]. However, to calculate the ground and finite-temperature states of many carriers strongly interacting with the phonon field turned out to be not easy task. Diagrammatic Monte-Carlo approach has allowed to solve it for the case of strong Holstein (short range) EPI [31] resulting in small (with the size about one unit cell) (bi)polaron formation [32,33]; no coexistence of small (bi)polarons with delocalized carriers was obtained. However, application of diagrammatic Monte-Carlo method to the case of many carriers and strong Fröhlich (long-range) EPI is hampered by the sign problem.

Strong Fröhlich EPI produces autolocalized carriers with the size much larger than the unit cell (large polarons and bipolarons) [20,24]. Bearing in mind strongly ionic lattice of cuprates and the fact that CDW period in them (the (bi)polaron size) is much larger than the lattice constant Fröhlich EPI seems to be more appropriate. For strong Fröhlich EPI in highly doped system a generalization of traditional for the polaron theory variational approach yields the in-plane size of the large bipolarons in the ground state [34]. It is somewhat smaller than the size of a single polaron or bipolaron formed in the same lattice but much larger than the in-plane lattice constant in cuprates. Then, according to the uncertainty relation, the maximum momentum $k_0$ of a carrier in a (bi)polaron is much smaller than the size of the first Brillouin zone (FBZ). The states with higher momentums are occupied by delocalized carriers (at sufficient carrier density). They do not screen EPI as to screen it delocalized carriers have to be localized inside the area occupied by autolocalized carriers that is not permitted by Pauli exclusion rule.

Thus, although many manifestations of small and large polarons and bipolarons are similar, (for example, broad bands in ARPES and optical spectra [21-26]) there is an important distinct of highly doped systems with strong Fröhlich EPI from ones with strong Holstein EPI. It is possibility of coexistence of autolocalized and delocalized carriers [35,34]. Sharing the momentum space between them regulated by Pauli exclusion rule displays itself in ARPES spectra as "vertical dispersion" or waterfalls emerging due to different energetic cost of the system relaxation after photoemission from autolocalized and delocalized states [36]. Such features are universally observed in ARPES spectra of both electron- and hole-doped cuprates as well as undoped parent compounds [37-41]. They, of course, demonstrate only presence of electron bipolarons near the band bottom, formation of hole bipolarons after the photoemission displays itself in absence of spectral weight near the FS in ARPES spectra [36] observed in undoped and hole-doped cuprates with doping $p<0.1$ [29,37,39]. Both CO wave vector $K_{CO}$ and the waterfall wave vector $k_0$ are related with the bipolaron size. As a result, $K_{CO}$ expressed in r.l.u. used in RXS ($2\pi/a$) is predicted to be close to $k_0$ in r.l.u. used in ARPES ($\pi/a$) [34], where $a$ is the lattice constant, and such coincidence is indeed observed in cuprates [3,37-41].

Once interpreting the unconventional CDW in cuprates as a manifestation of large-(bi)polaron liquid the other phases (PG and SC) turn out to be naturally related with its presence and coexistence with delocalized carriers. Antinodal PG in ARPES and STS spectra of the two-liquid system appears due to impact of the potential created by autolocalized carries on the states of delocalized carriers. It transforms Bloch quasiparticles into distributed in space wave packets, in which the carrier momentum is various in areas with various additional potential [42]. Topology of the hole-like dispersion and flat dispersion near antinodes makes impossible the propagation of near-antinodal distributed wave packets in areas with high negative additional potential energy emerging due to (bi)polarons' potential [42]. As a result, density of states near FS observed in STS experiments decreases and ARPES spectrum intensity near antinodes falls near the FS. Predicted values of the PG width and of the so-called FS angle of the PG closing as well as obtained Fermi momentum mismatch [42] (particle-hole asymmetry) are in consent with experiments on cuprates [4,5]. SC phase in the two-liquid system occurs below the temperature of the superfluid transition in the bipolaron liquid.

Below we consider what area do the regions in which PG, CDW and SC can be observed occupy in the phase diagram of strongly interacting electron-phonon systems and compare the result with the phase diagram of hole-doped cuprates. To do this we minimize a free energy of the two-liquid system comprising a liquid of large polarons and bipolarons and Fermi-liquid of delocalized carriers at each fixed temperature varying the (bi)polaron radius. The former liquid is like large-bipolaron liquid considered by Emin [43], but with bipolarons of both signs of charge and higher density of carriers typical of cuprates, so that the interbipolaron distance may be close to their size. (Bi)polaron liquid is formed due to weak inter(bi)polaron interaction screened with static dielectric constant which is high in cuprates [44,45], short coherence length of the CDW characteristic of cuprates is natural for areas of local order in a liquid.

Comparison of free energy of the two-liquid system with the free energy of the same system without (bi)polarons determines the area of existence of the two-liquid system of carriers where the PG can be observed (provided the dispersion is hole-like). The region in the phase diagram in which CO is observed is somewhat smaller since to observe CO in experiments some minimal number of droplets of the (bi)polaron liquid is needed. To calculate the temperature of the superfluid transition in the large-bipolaron liquid, we apply a standard method of Landau Bose-liquid theory [46] and the excitation spectrum of the large-bipolaron liquid [43] for systems with one and two conducting layers in the unit cell. We use the carrier dispersion modelling the dispersion in the lower Hubbard band of cuprates which is formed due to carrier correlations. Thus, the strong electron correlations occurring in cuprates are taken into account primarily, before the EPI, as their energy scale is much larger. Coexistence of autolocalized and delocalized carriers results in the other, more subtle (occurring on a smaller energy scale) carrier correlations mediated by the crystal lattice, which are considered below.

It should be noted that BCS pairing and pairing at BCS-BEC crossover [47] differ essentially from the formation of bipolarons. The latter are formed by carriers with the energy near its minimum whereas BCS pairs and crossover pairs are formed by carriers in the vicinity of the FS [46]. The adiabatic condition for the BCS and BCS-BEC crossover cases is Fermi energy much higher than the phonon one [47], whereas for the autolocalized carrier the kinetic energy of the carrier motion in the polarization potential well should be much higher than the phonon energy [20]. The EPI constant used in BCS and crossover theories being proportional to the density of states near the FS is not large in cuprates. However, the EPI energy in strongly interacting electron and phonon fields (i. e. in systems where autolocalized carrier states are formed) does not depend on the density of states near the FS, it is determined by static and high-frequency dielectric constants [20,24]. At their values measured in cuprates [44,45,48] the calculated phase diagram demonstrates presence of two-liquid system of charge carriers and related areas with PG, CO and SC.

The article is organized as follows. The second part contains three subparts. In the first we briefly describe the distribution function for charge carriers in systems where autolocalized and delocalized carriers can coexist [35,49]. Then we discuss the carrier dispersion which models the dispersion in hole-doped cuprates and apply it to calculate the free energy increment caused by the PG opening. The last subpart recalls calculation of the bipolaron binding energy as function of its radius [34] and discusses the contributions to the free energy to be minimized. The third part considers the excitation spectrum of the bipolaron liquid with one and two conducting layers in the unit cell and its application to the calculation of the maximum bipolaron density in Bose-vapor at given temperature. Comparison of this density with the total density of hole bipolarons (at given doping and temperature) obtained from the distribution function determines the area where the SC phase exists. The fourth part presents the results of the calculations and their discussion, then there are conclusion and outlook.

## II. MODELS AND METHODS TO OBTAIN THE FREE ENERGY OF A TWO-LIQUID SYSTEM OF CHARGE CARRIERS WITH CUPRATES-LIKE DISPERSION.

Two-component ground normal state comprising autolocalized and delocalized carriers was obtained earlier in a system with strong Fröhlich EPI and high carrier density in the frames of the generalized variational approach [34]. Its evolution with the temperature is considered below to describe the phase diagram of such a system by determining the density of (bi)polaron liquid as a function of doping and temperature. To obtain it we minimize the system free energy at fixed temperature and doping varying the bipolaron size (related with the CDW period) and compare the obtained free energy with the free energy of the purely Fermi-liquid system.

### II.1. Distribution of carriers over autolocalized and delocalized states at hole doping

To calculate the free energy of a system in which coexistence of autolocalized and delocalized carriers is possible one needs distribution function for carriers over autolocalized and delocalized states. Obviously, a distribution over carrier states with certain momentum cannot be applied since the momentum of an autolocalized carrier has an uncertainty of the order of the momentum itself. A suitable distribution was constructed earlier [35,49] with Hibbs method. Here it is applied for a system with cuprates-like dispersion, therefore the integrals in the distribution function are written for the case of 2D momentum space since in cuprates the carrier mobility in the direction perpendicular to the conducting plane is negligible. However, the carrier wave function in the (bi)polaron is essentially 3D since according to the experimental data the lattice deformation in the CDW phase of cuprates occurs not only in the conducting plane but in the whole volume [12].

The normalizing condition for the distribution is written for a subsystem having an area in the coordinate space equal to the area $S_0$ of intersection of the (bi)polaron with the conducting plane (it is shown in Fig.1(a) with black and red dashed lines for electron and hole (bi)polarons, respectively) and an area in the momentum space whose size $k_0$ is determined by the (bi)polaron size through the uncertainty relation as is described below. STM and RXS experiments reveal tetragonal coordination of the CO in the conducting plane of cuprates [3,50,51], likely due to the symmetry of the lattice. Therefore we use square area per one (bi)polaron in the coordinate space with the side $2R_{bip}$ (it is concurrently the CDW wavelength), so that $S_0 = 4R_{bip}^2$. Pauli exclusion rule prevents interbipolaron penetration of autolocalized electrons/holes across black/red dashed lines. In the free energy calculation below we suppose that 90% of the (bi)polaron charge is concentrated in the 90° rhombic areas with diagonals $2R_{bip}$ which are shown in Fig.1 as grey (blue) rhombs with black solid boundaries for an electron (hole) bipolaron.

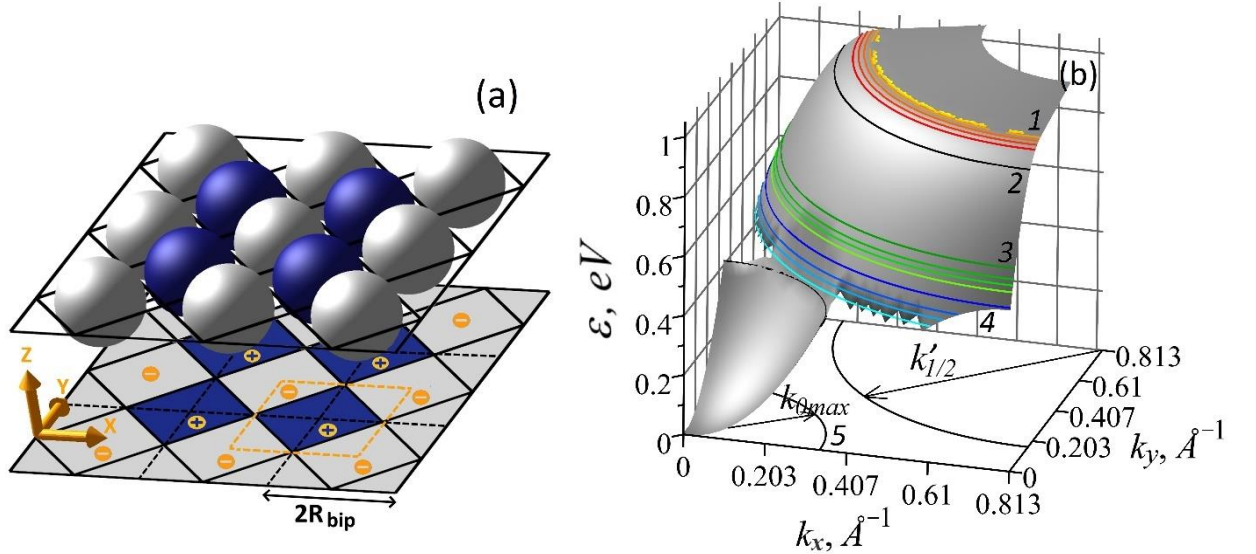

FIG.1 (a) Sketch of an area of the bipolaron liquid with the local order in a conducting plane (shown with black frame). Electron and hole bipolarons are depicted as differently colored spheres. Area per one electron (hole) bipolaron is shown with dashed black (red) lines in the bottom. Intersections of the conducting plane with regions containing 90% of charge of an electron (hole) bipolaron (90° rhombs with diagonals $2R_{bip}$) are highlighted with grey (blue) color in the bottom. (b) Carrier dispersion in the lower Hubbard band used in the calculations, parabolic near the band minimum and quasi-parabolic as is determined by Eq.(3) near the minimal hole energy (shown with line 1). Line 1 is the arc with the radius $k'_{min}$ determined by Eq.(5), shown for the minimum bipolaron radius $R_{bip}^{min} = 7$Å. Arcs (from orange to red) with slightly larger radii demonstrate positions of the minimal hole energy in systems with bipolaron radius $R_{bip} = 7.5, 8, 9, 10, 11, 12$Å as is described in II.2. Line 2 (the arc with the radius $k'_{1/2}$) represents the FS at zero doping in absence of bipolarons and PG. Lines 3 show the boundaries between the cold and hot hole states for the same set of $R_{bip}$ from 7 to 12 Å. FSs at doping $p=0.25$ for the same set of $R_{bip}$ are depicted with lines 4. For each set of lines a lighter color corresponds to a smaller $R_{bip}$. Line 5 shows the carrier energy at the maximum momentum $k_{0max}$ of cold electrons at $1/\varepsilon^*=0.3$, $\varepsilon_0=30$, $m_\parallel^*=m_e$, $m_\perp^*=40m_e$. Dispersion used in the free energy calculation is limited by line 4 for holes and line 5 for electrons. The maximum electron energy (at $k' = k'_{min}$) is chosen to connect the electron- and hole-like parts of the dispersion smoothly.

To simplify the integration in the distribution function we use isotropic dispersion for electrons and holes near their minimum energy. The maximum momentum $k_0$ of a carrier in the bipolaron is estimated as follows: we write the product of the uncertainty relations for $x$ and $y$ coordinate and projections of the momentum of a carrier in the bipolaron and then substitute the square area $\Delta k_x \Delta k_y$ with the circle area $\pi k_0^2$

$$4R_{bip}^2 \pi (\hbar k_0)^2 = (2\pi\hbar)^2, \quad k_0 = \frac{\sqrt{\pi}}{R_{bip}}. \tag{1}$$

Distribution of carriers of both signs of charge is needed to obtain the free energy of the two-liquid system of carriers, as it will be seen below. The size of the electron and hole polarons and bipolarons in the coordinate space is the same, therefore, the uncertainty of the hole momentum modulus in the hole (bi)polaron is also estimated as $k_0$. However, the hole momentum is counted from the arc-like FS (for a dispersion used below it consists of four arcs of the circumferences with the centers in $(\pm\pi, \pm\pi)$ points of the FBZ and the radius $k'_{min}$, one of them is shown with line 1 in Fig.1(b)) along the arc radius, and the area of the subsystem in the momentum space for the holes distribution is the area of a ring of the $k_0$ thick.

All the variants of filling the subsystem with zero, one or two carriers (with the opposite spins) which may be partially in autolocalized or delocalized states are taken into account in the normalizing condition [49]:

$$e^{\frac{\Omega}{kT}}\left(1 + e^{\frac{\mu}{kT}} \int_0^1 dN[NJ_1 + (1-N)J_2] + e^{\frac{2\mu}{kT}} \int_0^2 dN_1 \int_0^{2-N_1}[\sum_{i=1}^2 N_i I_i + (2 - N_1 - N_2)I_3]dN_2\right) = 1.$$

It yields the following partition function and density of bipolarons $n_{bip}$, polarons $n_1 \equiv n_{pol}$ and delocalized cold (having momentums $k<k_0$) carriers $n_2$ in a conducting plane:

$$Z = 1 + \frac{1}{2}e^{\frac{\mu}{kT}}\sum_{i=1}^2 J_i + \frac{4}{3}e^{\frac{2\mu}{kT}}\sum_{i=1}^3 I_i,$$

$$n_{bip} = \frac{4}{3}e^{\frac{2\mu}{kT}}I_3(ZS_0)^{-1}, \tag{2}$$

$$n_i = \left\{\frac{1}{2}e^{\frac{\mu}{kT}}J_i + \frac{8}{3}e^{\frac{2\mu}{kT}}I_i\right\}(ZS_0)^{-1},$$

$$J_i = \frac{1}{S_k}\int_{p_i^{min}}^{p_i^{max}} e^{-\frac{E_i(p)}{kT}} p\, dp, \quad I_1 = e^{-\frac{E_C}{kT}}J_1^2, \quad I_2 = J_2^2, \quad I_3 = J_3^2,$$

where $\mu$ is the chemical potential, $E_1 = E_{pol} + \varepsilon(p)M_{pol}^*/m^*$, $E_2 = \varepsilon(p)$, $E_3 = (E_{bip} + \varepsilon(p)M_{bip}^*/m^*)/2$, $p_{1,3}^{min} = 0$, $p_{1,3}^{max} = p_2^{min} = m^*u$, $p_2^{max} = \hbar k_0$; $\varepsilon(p)$ is the carrier dispersion with respect to its minimal energy, for electrons $\varepsilon(p) = p^2/2m^*$, for holes it will be defined below as well as $k'_{min}$ value; $u$ is the maximum group velocity of phonons strongly interacting with charge carriers; $E_{pol}$, $E_{bip}$, $M_{pol}^*$, $M_{bip}^*$ are binding energies and effective masses of the polaron and bipolaron, respectively, $E_C$ is the energy of Coulomb interaction between two polarons occupying the same area; $S_0$ is the area of the subsystem in the coordinate space, $S_k$ is the area of the subsystem in the momentum space, it is different for distribution of electrons: $S_k = \pi(\hbar k_0)^2$ and holes: $S_k = \pi \hbar^2 [(k'_{min} + k_0)^2 - k'^2_{min}]$.

Integration in Eqs.(2) is over the average carrier momentum as the instantaneous momentum does not have certain value in the autolocalized states. The integration limits in the distribution (2) take into account that the velocity of a large polaron and bipolaron is limited by the maximum group velocity $u$ of phonons participating in their formation (since they represent coupled wave packets of the carrier and polarization fields) [52,15]. Filling the subsystem with two carriers one of which is in the polaron and the other in a delocalized cold state is not possible due to necessity of the opposite spins of these carriers. If this were to be realized, then delocalized cold carriers should be localized in regions where polarons with the appropriate spin direction are present. In the bipolaron density $n_{bip}$, the factor 2, which comes from $-\partial\Omega/\partial\mu$, cancels out with 1/2 caused by the fact that the number of bipolarons is two times less than the number of carriers in the bipolaron state.

Dividing the area of the momentum space $k<k_0$ between autolocalized and delocalized carrier states does not limit filling of delocalized states with momentums $k>k_0$ which can be called hot carrier states. Their occupation is controlled by Fermi distribution. If the polaron and/or bipolaron states are completely filled the rest carriers can occupy only hot states. Equation for the carrier chemical potential demands equality between the sum of the carrier density in cold and hot states and the total carrier density.

## II.2. Modelling the dispersion of hole-doped cuprates and calculating PG-induced changes in the free energy

Fig.1(b) represents carrier dispersion used in the present consideration, it is aimed to model the carrier dispersion in the hole-doped cuprates and to simplify the free energy calculation. The minimal energy of electrons is supposed to be in Γ point of the FBZ as it occurs in cuprates. For the holes whose energy is constant approximately on four arcs in the FBZ one of which is shown in the bottom of Fig.1(b) we use the following approximation of the hole dispersion near the minimum of the hole energy:

$$\varepsilon(k') = c(k' - k'_{min})^2, \tag{3}$$

where $k'$ is the radius of a constant-energy curves which are the arcs of circumferences with the centers in $(\pm\pi/a, \pm\pi/a)$, $k'_{min}$ is the radius of the arc corresponding to minimal hole energy whose position will be determined below, and $c$ is a constant parameter.

Dispersion near the minimum of an electron energy can be expressed in the form $3.81k^2/m_\parallel^{*el}$ (eV), where $k$ is in Å$^{-1}$, i.e. for dispersion with a minimum in Γ-point $c=3.81$ corresponds to unit in-pane effective mass $1/m_{xx}^{*el} = 1/m_{yy}^{*el} \equiv 1/m_\parallel^{*el} = 1/m_e$. But for dispersion (3) with minimums on a ring and $c=3.81$ a radial in-plane component of the hole effective mass tensor near the minimum hole energy is unit whereas a tangential in-plane component is infinity. This would result in approximately two times lower expectation value of the carrier kinetic energy in (bi)polaron in comparison with minimum in Γ-point. In cartesian coordinates, used below in the (bi)polaron binding energy calculation such decrease in the expectation value of the carrier kinetic energy in (bi)polaron corresponds to $m_{xx}^{*h} = m_{yy}^{*h} = m_\parallel^{*h} = 2m_e$. However, the carrier effective mass in cuprates is about unity. Therefore we increase the dispersion steepness by using in (3) $c=3.81*1.6667$ (for energy $\varepsilon$ in eV and $k'$ in Å$^{-1}$) that allows using the in-plane hole effective mass $m_{xx}^{*h} = m_{yy}^{*h} = m_\parallel^{*h} = 1.2m_e$.

Fig.1(b) represents the electron dispersion, parabolic near the band bottom and determined by Eq. (3) in the area of hole-like dispersion. The value of the electron energy on the constant energy arc with the radius $k'_{min}$ is chosen in such a way to provide a smooth connection between the electron and hole dispersions in the nodal direction. This value is not included in the calculations, so it can be adapted to the hole dispersion with any $k'_{min}$. Energies of doped holes as well as the energies of electrons which are considered during variation of the free energy of the system (as it will be described below) do not go beyond the depicted parts of their parabolic (for electrons) and quasi-parabolic (for holes) dispersion. Rather flat character of the near-antinodal dispersion occurring in cuprates [29,9] is taken into account as is described below the Eq.(7). The hole dispersion with another position of the minimum is easily approximated in the region considered in the free energy variation with quasi-parabolic dispersion (3), as is exemplified in Fig.2(a,b). Deviation of one dispersion from the other is much smaller than the carrier kinetic energy in the (bi)polaron which coincides with the binding energy depicted in Fig.4 at large (bi)polaron radius and essentially exceeds the maximum binding energy at lower (bi)polaron radius.

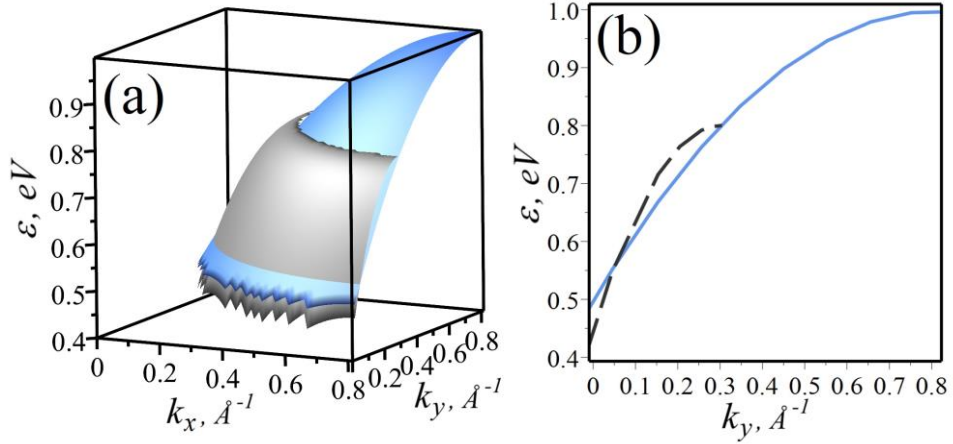

FIG.2. (a) An example of approximation of a dispersion with the maximums in $(\pm\pi/a, \pm\pi/a)$ FBZ points (blue surface) with the dispersion (3) (grey surface); (b) cross section of the dispersions represented in panel (a) by the plane $k_x = \pi/a$.

A factor strongly influencing free energy of the system under study is exclusion of near-antinodal delocalized states of the carrier in presence of bipolarons which displays itself in ARPES and STS spectra as a PG [42]. In more details, stationary states of delocalized carriers in additional potential of bipolarons change and can be described as distributed in space "wave packets" of Bloch waves in which the carrier momentum varies upon the quasiparticle (QP) propagation in the CO potential [42,53]. The QP momentum $\boldsymbol{k}_j$ in $j$-th cell of a square mesh in which the CO potential is approximated by a constant $U_j$ is determined from the system of equations

$$E(\boldsymbol{k}_{av}) = \varepsilon(\boldsymbol{k}_j) + U_j, \qquad k_{g,j-1} = k_{g,j}, \tag{4}$$

where $\boldsymbol{k}_{av}$ is the carrier momentum in areas with zero additional potential (called below the average momentum), so that $E(\boldsymbol{k}_{av}) = \varepsilon(\boldsymbol{k}_{av})$; $\varepsilon(\boldsymbol{k})$ is the dispersion of the Bloch carrier, and $g = x$ or $y$ depending on the orientation of the boundary between the $j-1$ and $j$ cells. An example of the region in which the momentum of a QP varies (a momentum trajectory) is shown in Fig.2(a) with light blue quadrangle. At using the dispersion (3) QP states can be indexed either by $\boldsymbol{k}_{av}$ or by the energy $E(\boldsymbol{k}_{av})$ and the angle $\alpha$ or $\varphi$ shown in Fig.2(a).

At hole-like dispersion the system of equations (4) does not have real solutions near antinodes (at $\varphi < \varphi_{PG}$) in areas with high negative additional potential energy [42,53]. The physical reason for this is flat hole dispersion near FS in antinodal region (large distances between the constant energy arcs corresponding to $\varepsilon(\boldsymbol{k}) = E$ and $\varepsilon(\boldsymbol{k}) = E + U_0$, where $U_0$ is the amplitude of the CO potential) and specific topology of the hole-like dispersion. They do not allow a QP propagating in near-antinodal directions to increase its "kinetic" energy $\varepsilon(\mathbf{k})$ enough to compensate for large negative potential energy $U$ and keeping their sum $E$ constant. In Fig.3(a) the momentum trajectory of existing QP is shown as blue quadrangle, a momentum trajectory of a state with a bit smaller φ and the same energy $E$ does not reach the arc $\varepsilon(\mathbf{k}) = E + U_0$ [42,53] because in areas with high negative additional potential energy the solutions of Eqs.(4) for the local carrier momentum in states with average carrier momentum near antinodes are complex. This results in extinction of these states so that they cease to be stationary.

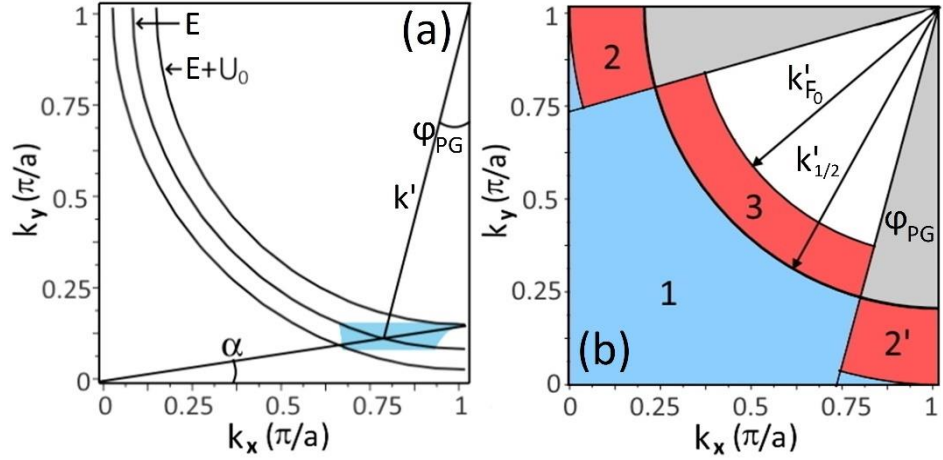

FIG.3 (a) Illustration of calculation of the angle $\varphi_{PG}$ bounding the area of the FBZ excluded due to the PG. Blue quadrangle depicts the area of the QP momentum trajectory, the arcs are curves where the Bloch carrier energy $\varepsilon = E(\mathbf{k}_{av})$ and $\varepsilon = E(\mathbf{k}_{av}) + U_0$, $\mathbf{k}_{av}$ is intersection of the constant energy $\varepsilon = E(\mathbf{k}_{av})$ curve and depicted ray from the origin, $U_0$ is the amplitude of the CO potential; (b) illustration of shift of the FS at zero doping due to the PG opening: the areas 2 and 2' becomes inaccessible, instead the area 3 becomes occupied by electrons.

To take into account change of the system free energy due to exclusion of near-antinodal QPs in presence of the CO ((bi)polaron) potential we calculate the angle $\varphi_{PG}$ bounding the excluded area as is shown in Fig.3(a,b). $\varphi_{PG}$ can be approximated as an angle coordinate of the intersection between the curve of the QP energy $E$ and a line containing the origin and the point where the arc corresponding to the energy $E+U_0$ meets the FBZ boundary as is illustrated by Fig.3(a) [42]. $\varphi_{PG}$ value depends on the carrier dispersion which determines the distance between the constant energy curves corresponding to the energies $E$ and $E+U_0$ and on the amplitude $U_0$ of the CO potential. The $U_0$ value is estimated as an amplitude of superposition of potentials created by several (8 in the present calculation) conducting layers with harmonic charge distribution and the total charge of the polaron and bipolaron $e/\varepsilon_0$ and $2e/\varepsilon_0$, respectively. The obtained $U_0$ value practically coincides with the potential in the center of a uniformly charged ball of the radius $R_{bip}$ with the total charge $e/\varepsilon_0$ and $2e/\varepsilon_0$, respectively.

Since the value of $\varphi_{PG}$ depends on $U_0$, the increment of the system energy due to the PG opening depends on the bipolaron radius. This increment can be easily calculated with the use of the dispersion (3). The PG opening results in absence of the carrier stationary states with average momentums near the antinodes (in the areas denoted as 2 and 2' in Fig.3(b)). Instead, the carriers occupy the area above the FS occurring in absence of the PG which is denoted as area 3 in the Fig.3(b). Calculating the areas 2 and 2' and equating it to the area 3 we obtain the position $k'_{min}$ of the FS at zero doping in presence of the PG (the PG-induced shift of the FS at zero doping with respect to its position in absence of bipolarons):

$$k'_{min} = \left(k'^2_{1/2} - [\left(\frac{\pi}{a}\right)^2 - k'^2_{1/2}]\frac{8\varphi_{PG}}{2\pi - 8\varphi_{PG}}\right)^{1/2}, \qquad (5)$$

where $k'_{1/2}$ is the radius of the arc representing the FS at zero doping in absence of the PG shown with line 2 in Fig.1(b). In cuprates, the FS shift may not be as pronounced due to the shape of the constant energy curves, which are better described as arcs centered outside the FBZ. Such "more rhombic" dispersion together with decrease of the $\varphi_{PG}$ at increasing hole energy [42] which is not taken into account in the present calculation result in decrease of the area of excluded regions 2 and 2' in Fig.3(b).

Since the value of $\varphi_{PG}$ depends on the bipolaron radius, $k'_{min}$ value determined by Eq.(5) also depends on it. The smaller the bipolaron radius the larger the $\varphi_{PG}$ value and the shift of the FS at zero doping, so that the smaller the $k'_{min}$ value. As calculation shows (Fig.6(b)) the bipolaron radius is constant at increasing temperature up to temperatures higher than $T_c$, then it increases. To prevent changing the carrier dispersion at change of the bipolaron radius with temperature we use in the dispersion (3) the minimal $k'_{min}$ value corresponding to the minimal (among occurring for given medium parameters) bipolaron radius $R_{bip}^{min}$. Energy, corresponding to this $k'_{min}$ is shown in Fig.1(b) with yellow line 1. Larger values of the bipolaron radius correspond to larger $k'_{min}$ and, hence, to a lower electron Fermi energy at zero doping shown in Fig.1(b) with lines colored from light orange to red (they lie just below line 1). Below we denote these larger $k'_{min}$ as $k'_{F0}$. Green lines 3 in Fig.1(b) show the minimal energy of hot holes $\varepsilon(k'_{F0} + k_0)$ for several values of the (bi)polaron radius, where $k_0$ is the maximum momentum of a hole in the (bi)polaron determined by Eq.(1).

Thus, entering the distribution (2) hole energy with respect to its minimal energy (corresponding to the momentum $k'_{F0}$) is described by Eq.(3) with shifted zero energy as is follows:
$$\varepsilon(k') = c(k' - k'_{min})^d - c(k'_{F0} - k'_{min})^d, \qquad (6)$$
Using Eqs. (5,6) one can calculate the increment of density of the system energy due to the PG opening as is illustrated by Fig.3(b):
$$\Delta E^{PG} = 2[(2\pi - 8\varphi_0)\int_{k'_{F0}}^{k'_{1/2}} \varepsilon(k')\,k'dk' - 8\varphi_0 \int_{k'_{1/2}}^{\frac{\pi}{a}} \varepsilon(k')\,k'dk']/(2\pi)^2, \qquad (7)$$
where 2 is due to spins and the upper limit in the integration $\pi/a$ is the boundary of the hole-like dispersion. To take into account rather flat near-antinodal dispersion characteristic of cuprates we substitute $\varepsilon(k')$ in the second term of Eq.(7) representing energies of the near-antinodal carriers with the constant $\varepsilon(k'_{1/2})$.

Due to different distances between the constant-energy arcs corresponding to the QP energy $E$ and $E+U_0$ (depicted in Fig.3(a)) at different $E$ the value of $\varphi_{PG}$ depends on the QP energy $E$ [45]. Including this dependence into the integrals of the distribution (2) increases dramatically the calculation time. Therefore here we use $\varphi_{PG}$ value calculated for an average energy $\bar{E}$ of a QP with the hole-like dispersion (averaged over the hole-like dispersion range of $k'$ from $k'_{min}$ to $\pi/a$). Since $k'_{min}$ value slightly depends on the minimum bipolaron radius $R_{bip}^{min}$, the average energy also slightly changes with $R_{bip}^{min}$. We calculated the phase diagram for several values of $\bar{E}$ corresponding to different $R_{bip}^{min}$ and found that it did not change noticeably.

### II.3. Free energy variation in a system with doped lower Hubbard band

When the carrier density is not low the size of the autolocalized carrier states (polaron and bipolaron) corresponding to the lowest free energy of the system and, consequently, their binding energy entering the distribution function (2) depend on the carrier density because Pauli exclusion rule prevents interpenetration of (bi)polarons of the same sign of charge. To obtain the bipolaron binding energy as function of its size we use conditional variation method [34]. It searches for the minimum of expectation value of Fröhlich bipolaron Hamiltonian in a state described by trial wave function of carriers in the bipolaron and a vector of a coherent state of the polarization (phonon) field [15] under the condition that the bipolaron size has a given fixed value. Fröhlich bipolaron Hamiltonian has the form

$$H = \sum_{i=1}^{2}\left(-\frac{\hbar^2}{2\widehat{m^*}}\Delta_{r_i} - \sum_{k}\frac{e}{k}\sqrt{\frac{2\pi\hbar\omega_k}{V\varepsilon^*}}[b_k e^{ikr_i} + b_k^+ e^{-ikr_i}]\right) + \sum_{k}\hbar\omega_k b_k^+ b_k + \frac{e^2}{\varepsilon_\infty}\frac{1}{|r_1-r_2|},\tag{8}$$

where $1/\widehat{m^*}$ is inverse effective mass tensor with strongly different in-plane $1/m^*_{xx} = 1/m^*_{yy} \equiv 1/m^*_\parallel$ and out-of-plane $1/m^*_\perp$ components ($m^*_\perp$ is considered $40m_e$), $1/\varepsilon^* = 1/\varepsilon_\infty - 1/\varepsilon_0$ is inverse effective dielectric constant, determined by the lattice polarizability. Here we characterize the in-plane bipolaron size with $R_{bip}$ which is a half of diagonal of 90° rhombic area (shown with blue/grey color in Fig.1(a)) enclosing 90% of the bipolaron charge (for short $R_{bip}$ is called the bipolaron radius):

$$\int_{V_{bip}}|\psi_{bip}(\mathbf{r_1},\mathbf{r_2})|^2 d^3r_1 d^3r_2 = 0.9. \tag{9}$$

In the perpendicular to the conducting plane direction (along z-axis) the bipolaron size cannot be considered as tending to zero since X-ray diffraction shows non-zero frozen shifts of all the ions in the unit cell, not only of those in the conducting plane [12]. Therefore the height (half of the height) of the unit cell was used as the bipolaron size along z axis for the case of one (two) conducting layer(s) in the unit cell. The polaron height is determined in the same way.

As a trial wave function of carriers in the bipolaron we use a normalized 3D generalization [34] of the 2D four-lobe wave function suggested by Emin [54]:

$$\psi_{bip} = \phi_{x1}\phi_{y2} + \phi_{x2}\phi_{y1}$$
$$\phi_{xi} = e^{-\frac{x_i^2+y_i^2}{2a^2}}e^{-\beta z_i^2}\cosh\left(\frac{lx_i}{a^2}\right) \tag{10}$$
$$\phi_{yi} = e^{-\frac{x_i^2+y_i^2}{2a^2}}e^{-\beta z_i^2}\cosh\left(\frac{ly_i}{a^2}\right),$$

where $a$ and $l$ are variational parameters. For the polaron the simplest trial function $\psi_{pol} = A\exp[-\alpha(x^2+y^2) - \beta z^2]$ was used. The value of $\beta$ was determined from the condition of 20 times decrease of the squared wave function of each carrier in the bipolaron and of the carrier in the polaron at $|z|=h/2$, where $h$ is the (bi)polaron thickness ($h$ used here is 13.2Å and 6Å or 4.1Å, for one and two conducting layers in the unit cell, respectively) that allows satisfying the condition (9). The polaron energy as function of its radius for given polaron thickness is obtained without minimization from the Fröhlich Hamiltonian and the condition $\int_{V_{pol}}|\psi_{pol}(\mathbf{r})|^2 d^3r = 0.9$. Adiabatic condition (the carrier kinetic energy much higher than the phonon energy) for contracted (bi)polaron is satisfied more readily than for non-contracted one as the energy of intra(bi)polaron motion of a carrier becomes higher at smaller (bi)polaron size according to the uncertainty relation.

Eqs. (2) and expressions for a spectrum of elementary excitations of the large-bipolaron liquid deduced below contain the polaron and bipolaron effective masses. Bipolaron effective mass is expressed similarly to the large-polaron effective mass [55,56] with replacement of Fourier-transform $\xi(\mathbf{k})$ of the squared modulus of the carrier wave function in the polaron by a sum of Fourier-transforms $\eta(\mathbf{k_1},\mathbf{k_2})$ of the squared modulus of the carriers wave function $\psi(\mathbf{r_1},\mathbf{r_2})$ in the bipolaron (10):

$$M^*_{pol} = \frac{e^2}{3\pi^2\varepsilon^*\Omega^2}\int \xi(\mathbf{k})^2 d^3k,$$
$$M^*_{bip} = \frac{(2e)^2}{3\pi^2\varepsilon^*\Omega^2}\int[\eta(\mathbf{k},0) + \eta(0,\mathbf{k})]^2 d^3k, \tag{11}$$

$$\eta(\boldsymbol{k_1}, \boldsymbol{k_2}) = \int |\psi(\boldsymbol{r_1},\boldsymbol{r_2})|^2 \exp(i\boldsymbol{k_1}\boldsymbol{r_1}) \exp(i\boldsymbol{k_2}\boldsymbol{r_2})\, d^3r_1 d^3r_2;$$

where 3 in the denominator corresponds to isotropic 3D case, in the 2D case it is substituted with 2, $\Omega$ is the average frequency of phonons interacting with the charge carrier, and the spatial dispersion of the lattice polarizability is neglected.

Examples of the obtained binding energy and effective mass of the polaron and bipolaron as functions of their radius are represented in Fig.4. It is interesting that effective mass of large (bi)polarons increases with the increase of the carrier density similarly to the small polaron mass behavior obtained with diagrammatic Monte-Carlo method [31]. Indeed, the equilibrium bipolaron radius at high carrier density is smaller than the radius of a single bipolaron [34], and, as is seen from Fig.4(c), the smaller the large bipolaron, the greater its effective mass. Taking into account spatial dispersion of the lattice polarizability (i.e. the ability of the polarization to propagate with the velocity not higher than the maximum group velocity of phonons $u$) may somewhat reduce the (bi)polaron effective mass [56]. It should be noted that effective masses (11) describe the independent motion of bipolarons during which autolocalized carriers drug their polarization coat. But at coherent superfluid motion of the bipolaron condensate, on the opposite, the polarization field propagates as a wave drugging the autolocalized carriers so that effective mass of bipolarons in condensate is much lower than (11) and may be close to double electron mass.

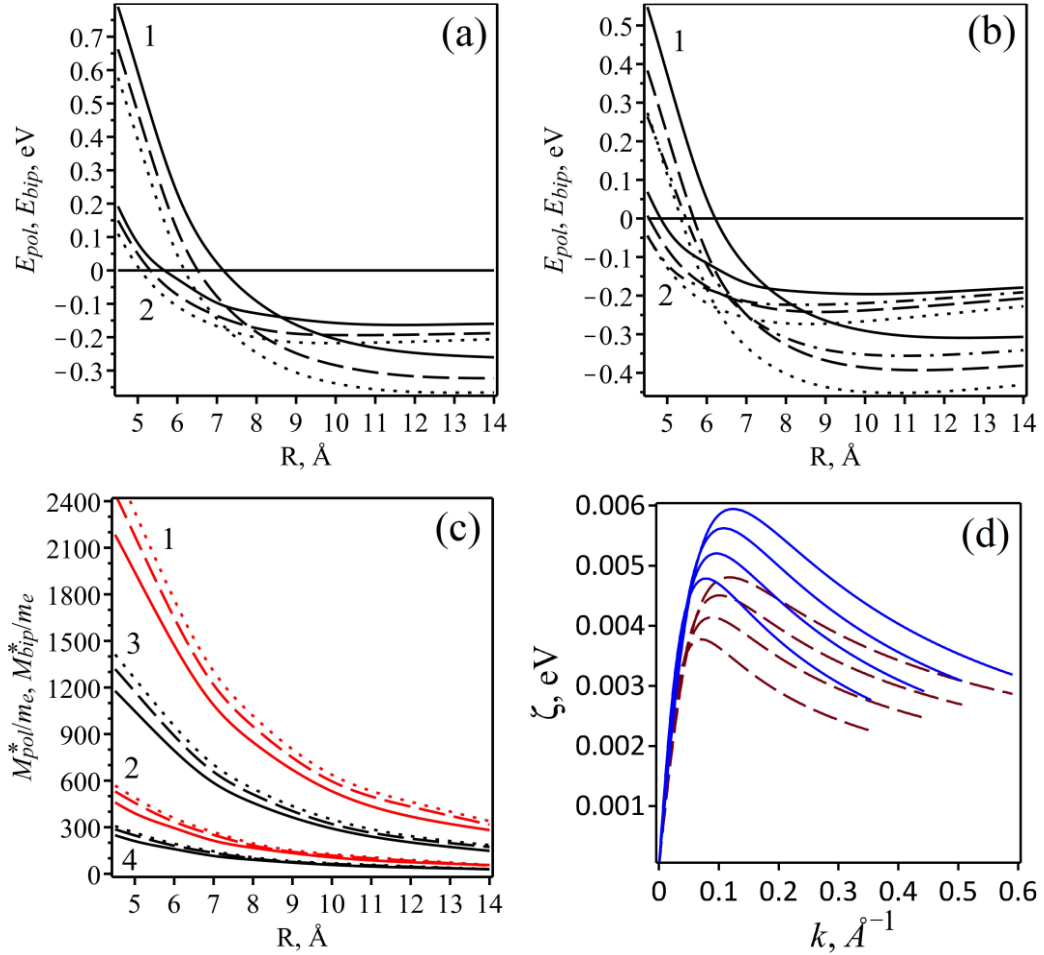

FIG.4(a) and (b) Binding energy of the bipolaron (1) and polaron (2) as functions of their radius in systems with 1 and 2 conducing planes in the unit cell, respectively; (c) effective mass of the bipolaron (1,3) and polaron (2,4) as functions of their radius in systems with 2 (1,2) and 1 (3,4) conducing planes in the unit cell. Solid lines: $1/\varepsilon^*=0.25$, $m_\parallel^*=m_e$; dashed lines: $1/\varepsilon^*=0.28$, $m_\parallel^*=m_e$, dotted lines: $1/\varepsilon^*=0.3$, $m_\parallel^*=m_e$, dashed-dotted lines: $1/\varepsilon^*=0.25$, $m_\parallel^*=1.2m_e$; $\varepsilon_0=30$ and $m_\perp^*=40m_e$ for all the lines, interplane distance for systems with two conducting planes in a unit cell is 4.1Å, the (bi)polaron thickness $h$=6Å and 13.6Å in 2- and 1-layer systems, respectively. (d) Spectrum of the elementary excitations of the bipolaron liquid for a system with 1 (dashed lines) and 2 (solid lines) conducting planes in the unit cell calculated according to Eqs.(17,18) for the bipolaron radius (from the bottom to the top) $R_{bip}$=6,7,8 and 10Å and interplane distance (for a system with 2 conducting planes in a unit cell) 4.1Å. End point for each dispersion is estimated according to Eq.(14).

Using in calculation of $E_{pol}$, $E_{bip}$, $M^*_{pol}$ and $M^*_{bip}$ the hole effective mass near the minimum of the hole energy is slightly inaccurate when the (bi)polaron radius departs from the minimum radius $R^{min}_{bip}$ for given medium parameters, as this results in a small shift of the zero-energy momentum from $k'_{min}$ to $k'_{F0}$. However, the inaccuracy is slight since shift of the zero-energy momentum apart of $k'_{min}$ is small as is illustrated by lines just below line 1 in Fig.1(b) and the shift of the minimum hole energy is small too because it occurs near the extremum.

Peculiar character of the Mott-Hubbard dielectrics results in location of the hole states with the minimum energy not in (π,π) and three equivalent points of the FBZ but on arcs dividing each quadrant of the FBZ into two equal areas shown in the bottom of Fig.1(b). Luttinger and Kohn [57] generalized the effective mass approach to systems with several minima of the carrier dispersion in the FBZ. Applying the Luttinger-Kohn result [57] to the case of infinite number of k-points corresponding to the minimal carrier energy (as it occurs in dispersion (3)) and using the dispersion (3) for holes allow similar calculating the binding energy of the (bi)polaron as function of its radius for hole and electron (bi)polarons.

Using the obtained binding energies of the polaron and bipolaron as functions of the radius we calculate the carrier density in all the states by minimizing the system free energy at given temperature, the variation parameter is the (bi)polaron radius. In neglect of the bipolaron interaction the free energy of the (bi)polaron liquid is additive due to (bi)polarons locality. Interaction between the (bi)polarons is taken into account in the mean field approximation, that preserves additivity of the free energy of the system. Potential created by bipolarons and polarons is estimated as the superposition of potentials of point charges ±2$e/\varepsilon_0$ and ±$e/\varepsilon_0$, respectively, located in the average positions of the (bi)polarons centers. When both polarons and bipolarons are present in the system, the potential created by them is estimated as linear combination of the two mentioned potentials with the weights proportional to relative densities of bipolarons and polarons $n_{bip}/(n_{bip}+n_{pol})$ and $n_{pol}/(n_{bip}+n_{pol})$.

Constructing the system free energy for minimization, it is necessary to take into account that the system under study is somewhat special. Since holes are considered on top of filled with electrons half of the lower Hubbard band, the free energy of the system under study varies depending on the fact whether the PG in the carrier spectrum is opened or not as is described by Eq.(7) and whether electron bipolarons are present in the system. Therefore, minimizing the system free energy at fixed temperature besides the energy of hole (bi)polarons and delocalized holes one should take into account the difference in energies between electron (bi)polarons and delocalized electrons in the same area of the momentum space near the band bottom. The radius of the area in the momentum space that may be occupied by autolocalized electrons (its boundary is shown with line 5 in Fig.1(b)) is the maximum value of $k_0$ (1) $k_{0max} = \frac{\sqrt{\pi}}{R^{min}_{bip}}$, where $R^{min}_{bip}$ is the minimum radius of the (bi)polaron at given medium parameters. Energy of the electrons with momentums higher than $k_{0max}$ need not be taken into account during minimization as it remains constant at the (bi)polaron radius variation.

Thus, we consider two distributions of the form (2) for the electrons and holes supposing electron and hole bipolarons have the same radius. The area of the electron bipolaron (of the region in which 90% of the electron density in the bipolaron is concentrated) intersection with the conducting plane is $2R_{bip}^2$ and the hole bipolaron area is also $2R_{bip}^2$, together they form an "elementary cell" of the local order in the bipolaron liquid with the area $4R_{bip}^2$. It is the minimal area whose free energy is needed to obtain the free energy density for subsequent minimization. The density of the free energy of the system to be minimized at a fixed temperature,

$$F = \frac{n_{bip}^h E_{bip}^h + n_{bip}^{el} E_{bip}^{el} + n_{pol\,i}^h E_{pol\,i}^h + n_{pol\,i}^{el} E_{pol\,i}^{el} + E_{colddel}^h + E_{colddel}^{el}}{4R_{bip}^2} + E_{hot}^h + E_{hot}^{el} + \Delta E_{PG} + E_{int}, \quad (12)$$

comprises the density of energy of cold and hot holes, the density of energy of cold and hot electrons with momentums up to $k_{0max}$, the increment $\Delta E_{PG}$ (7) in the electron energy density in presence of the PG and the density of energy of inter(bi)polaron interaction $E_{int}$. Contributions to the free energy from the first and second terms in the density $n_{pol}$ of polarons (2) ($n_{pol\,1}$ and $n_{pol\,2}$) are different due to energy $E_C$ of Coulomb repulsion of two polarons in the same area: $E_{pol\,1} = E_{pol}$ and $E_{pol\,2} = E_{pol} + E_C/2$, respectively. Kinetic energy of (bi)polarons is neglected in (12) due to their large effective mass and low maximum velocities. The density of energy of hot electrons with momenta up to $k_{0max}$ and of hot holes is obtained using their dispersion and Fermi distribution. The energy of delocalized cold carriers in the area $4R_{bip}^2$ is calculated as a product of their number in this area by the average energy of carriers with momenta from 0 to $k_0$ for electrons and from $k'_{F0}$ to $k'_{F0} + k_0$ for holes.

The density of free energy (12) of the system with (bi)polarons is minimized over the (bi)polaron radius $R_{bip}$. The minimization is carried out twice: first with the minimum possible value of $R_{bip}^{min}$ estimated from the (bi)polaron energy as function of its radius and then with the $R_{bip}^{min}$ obtained in the first minimization (if it differs significantly). The obtained minimum value of (12) at given doping $p$ and temperature $T$ is compared with the density of free energy of the system comprising all the doped holes and the electrons with momentums up to $k_{0max}$ with the same dispersion but without (bi)polarons at the same $p$ and $T$. In principle, the free energy of the system without (bi)polarons should be calculated with taken into account EPI. However EPI influences delocalized carrier states only for carriers in the vicinity of FS so that its influence on the free energy of the system is quite small.

In the region of phase diagram where the bipolaron liquid exists the obtained bipolaron density $n_{bip}$ and radius $R_{bip}$ as functions of doping and temperature are used to calculate the temperature of the SC transition $T_c$ as is described below and the charge ordering wave vector $K_{CO}$. The PG critical temperature $T^*$ is estimated as the lower of two temperatures – the temperature at which the energy of the system with (bi)polarons is equal to the energy of the system without (bi)polarons for given doping, and the temperature at which 5% of the maximum (for given parameters of the medium) density of (bi)polarons is present.

The CO onset temperature $T_{CO}$ is somewhat lower than $T^*$ as some minimal quantity of droplets of the (bi)polaron liquid is needed to observe the CO maximum in spectra of resonance X-ray scattering (RXS) and scanning tunneling spectroscopy (STS). The density of (bi)polarons necessary to observe the CO caused by (bi)polaron liquid depends also on the resolution of the method and equipment used. Therefore, it is not possible to point out the precise boundary of the CO region in the phase diagram constructed. We estimate $T_{CO}$ as the lowest of two temperatures: the temperature corresponding to presence of approximately 30% of the maximum number of hole (bi)polarons at given medium parameters and $0.7T^*$. The duality in determining $T^*$ and $T_{CO}$ is caused by imperfectness of the model used. Namely, the (bi)polaron number in the area $4R_{bip}^2$ at given temperature is determined by their binding energy and temperature according to the distribution (2). Ideally it would be determined by difference between the free energies of the phase with (bi)polarons and without them.

# III. TEMPERATURE OF THE SUPERFLUID TRANSITION IN LARGE-BIPOLARON LIQUID

To deduce the temperature of Bose-condensation of the bipolaron liquid we apply a standard method of Bose-liquid theory [46]. First it calculates the momentum $\boldsymbol{P}_n$ (per unit area of a conducting layer in the considered quasi-2D system) of the normal part of the Bose-liquid at small velocity $\boldsymbol{v}$ of the liquid. The momentum depends on the spectrum of elementary excitations $\zeta(\mathbf{k})$ of Bose-liquid and temperature [46]:

$$\boldsymbol{P}_n = \int \hbar \boldsymbol{k} n(\zeta - \hbar(\boldsymbol{k}\boldsymbol{v})) \frac{d^2k}{(2\pi)^2} \approx -\frac{\boldsymbol{v}\hbar^2}{2} \int k^2 \frac{\partial n}{\partial \zeta} \frac{d^2k}{(2\pi)^2}, \tag{13}$$

where expansion of Bose distribution $n(\zeta - \boldsymbol{p}\boldsymbol{v})$ correct at small $\boldsymbol{v}$ is used and an isotropy in the conducting plane is supposed that allows to write $\boldsymbol{P}_n = M\boldsymbol{v}$. We estimate the limit of the integration in Eq.(13) using the conservation of the number of degrees of freedom. Let the area of the integration in isotropic momentum space is a circle with the radius $k_{max}$. The number of states in this area is $N = S\pi k_{max}^2/(2\pi)^2$, where $S$ is the area of the crystal. $N$ should be equal to the number of degrees of freedom of bipolarons in the area $S$: $N = 2S * 2/(4R_{bip}^2)$, where the first 2 is the number of degrees of freedom for one bipolaron in the conducting plane and the second 2 is the number of bipolarons in the area $4R_{bip}^2$. Equating, one obtains

$$k_{max} = 2\sqrt{\pi}/R_{bip}. \tag{14}$$

The ratio of the momentum (13) to the liquid velocity $\boldsymbol{v}$ represents the mass of normal component of the liquid (per unit area of one conducting layer). Dividing it by the bipolaron effective mass $M_{bip}^*$ one obtains the density of the bipolarons in Bose-vapor in one conducting layer at given temperature:

$$n_{vap} = \frac{\hbar^2}{4\pi M_{bip}^* k_B T} \int_0^{k_{max}} \frac{e^{\zeta/k_B T}}{\left(e^{\zeta/k_B T}-1\right)^2} k^3 dk. \tag{15}$$

If the density of bipolarons $n_{bip}$ in the conducting layer exceeds their density in the Bose-vapor (15), the excess bipolarons condense. Thus, comparing the bipolarons density in the Bose-vapor (15) at given temperature with their density in the system (obtained by minimizing the system free energy) at the same temperature we find the temperature of the superfluid transition. The superfluid density determining the phase stiffness which is measured experimentally [58] is obtained as the difference between the total bipolaron density and their density in the Bose-vapor (15). The result is compared with the experimental data [58] in the Results section.

To calculate the bipolaron density in the Bose-vapor (15), one needs spectrum of the elementary excitations of the large-bipolaron liquid. It was obtained earlier for bipolarons of one sign of charge [43] with a standard method of Bose-liquid theory [46]. The method expresses the spectrum of the low-energy excitations of Bose-liquid through Fourier-transform of the interparticle interaction provided it is weak. This is the case for large bipolarons whose interaction is screened with static dielectric constant which is high in cuprates [44,45]. The interactions considered in [43] are long-range Coulomb interaction of bipolarons (of one sign of charge) and short-range repulsion of carriers in different bipolarons caused by Pauli exclusion rule. However, in the system under study there are bipolarons of both signs of charge; calculations show that the difference in the density of electron and hole bipolarons significantly increases the free energy of the system, so that they practically coincide. Taking this into account results in compensation of repulsive and attractive Coulomb interactions of bipolarons. Then only the term describing short-range repulsion of bipolarons of the same sign of charge [43] remains:

$$I(s) = e^2 \sum_n e^{-s/r_{\min}(n)}/(\varepsilon_\infty r_{min}(n)), \tag{16}$$

where $s$ is the distance between bipolarons and $n$ is the number of interplane distances between them. For the short-range interaction in one conducting layer the sum contains only $n=0$ term [43] and $r_{min} = 2R_{bip}$. Then the spectrum of elementary excitations of the large-bipolaron liquid in a system with one conducting layer in the unit cell has the following form [43]:

$$\zeta_1(k) = \frac{\hbar w_s k R_{bip}}{[1+(2kR_{bip})^2]^{3/4}}, \quad \hbar w_s = \sqrt{\frac{4\pi e^2 n_{bip} \hbar^2}{\varepsilon_\infty M_{bip}^* R_{bip}}} \quad (17)$$

where $n_{bip}$ is the in-plane density of bipolarons (2).

In [43] the short-range repulsion of carriers was considered only in one conducting layer. To calculate the temperature of the superconducting transition in a system with two conducting layers in the unit cell we take into account the short-range repulsion of bipolarons in neighboring layers in the same form as intralayer repulsion but with $r_{min} = \sqrt{d^2 + 2R_{bip}^2}$, where $d$ is a distance between conducting planes in a unit cell and $R_{bip}\sqrt{2}$ is the in-plane distance to the bipolaron with the opposite charge (in the neighboring layer the bipolaron charges are in the antiphase due to Coulomb interaction as is observed in cuprates [59]). Fourier-transform of the interlayer interaction is calculated by differentiating the integral 6.616(2) [60] over the parameter. Resulting spectrum has the following form:

$$\zeta_2(k) = \hbar w_s \sqrt{\frac{(kR_{bip})^2}{[1+(2kR_{bip})^2]^{3/2}} + \frac{k^2 d^3 R_{bip}}{4 r_{min}^2} \frac{e^{-s}(1+s)}{s^3}}, \quad s = \sqrt{k^2 d^2 + d^2/r_{min}^2}. \quad (18)$$

Fig.4(d) represents examples of the low-energy excitation spectra (17,18) of the bipolaron liquid, reminiscent of phason excitations of the CDW [61]. Of course, the spectrum (17,18) is only the first approximation intended to describe the lowest-energy excitations which mainly determine the temperature of the superfluid transition. More sophisticated consideration can refine or augment it, for example, with some roton-like excitations like that occurring in liquid $He^4$. It is important to test if a gap may appear in the spectra (17,18) in presence of defects or impurities as it occurs for conventional CDW since in such a case the temperature of the superconducting transition will increase.

## V. RESULTS AND DISCUSSION

The calculation shows that in a wide range of the medium parameters there exists an area in the phase diagram in which a two-liquid state of the carrier system comprising (bi)polaron and Fermi liquid has free energy lower than the state without (bi)polaron liquid. Then there are three physically different areas in the phase diagram of such systems in the interval of doping where the SC exists. They are the area without (bi)polaron liquid (with only Fermi liquid of delocalized carriers present), the area with presence of normal (bi)polaron liquid and Fermi liquid, and the area where both normal and superfluid components of the (bi)polaron liquid together with the Fermi liquid of delocalized carriers exist.

The area in the phase diagram where the PG is observed coincides with the region of existence of the (bi)polaron liquid with the only exclusion. When upon the increase of doping at some $p=p^*$ the FS becomes electron-like, the PG ceases to be observable [42], though the carrier states in the areas 2 and 2' in Fig.3(b) are still absent (as long as (bi)polarons are present in the system). The value of critical doping $p^*$ corresponding to disappearance of PG in ARPES and STM spectra due to transition from the hole-like to electron-like FS depends in the present approach on $R_{bip}^{min}$ as it influences the momentum $k'_{min}$ corresponding to the minimum hole energy. For example, calculation of FSs (like ones represented at Fig.1(b)) for different doping shows that at $R_{bip}^{min} = 7$Å $p^* = 0.19$ which is typical for cuprates. Thus, in the considered approach CO may occur in absence of the PG at $p > p^*$ as was observed in cuprates [9]. The superconducting phase exists in the area of the phase diagram in which the superfluid component of the bipolaron liquid exists, its position is calculated as is described in Section III.

Fig.5 represents the calculated position of the regions in the phase diagram where PG, CO and SC occur for systems with one and two conducting layers in the unit cell. Position of these areas is in agreement with their position in phase diagram of hole-doped cuprates [7, 3], sketch of which for YBaCuO system having two conducting layers in the unit cell is shown in Fig.5(e). As is seen from comparison of panels (a-c) and (d) the superconducting transition temperature $T_c$ essentially increases in a system with two conducting layers in the unit cell (panel (d)) in comparison with systems with one conducting layer (a-c) like it occurs in cuprates. The reason is higher energy of the elementary excitations of the bipolaron liquid in a two-layer system as is seen from comparison of Eqs. (17) and (18) and higher density of bipolarons in a two-layer system (at contraction along z axis the binding energy of bipolarons increases that results in a smaller equilibrium in-plane size).

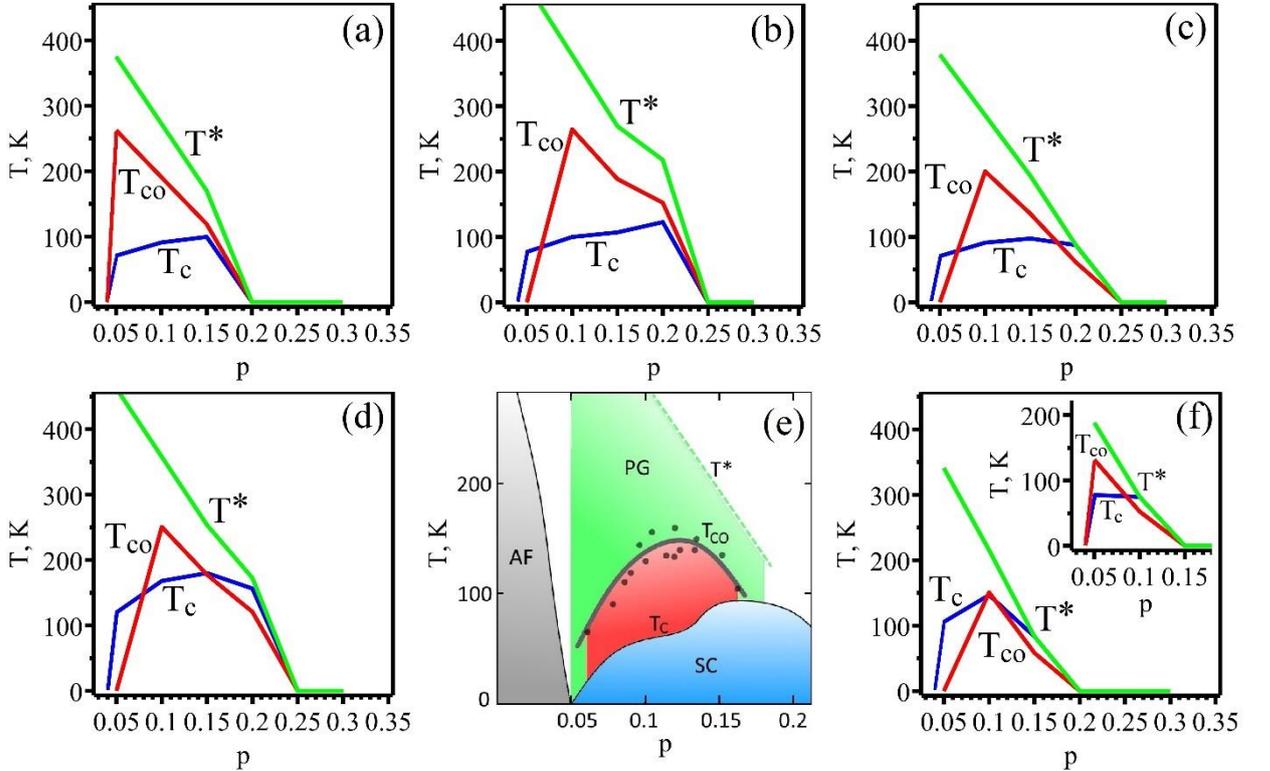

FIG.5(a-d,f) Calculated phase diagrams (below green line $T^*$ the two-liquid phase exists and PG can be observed at hole-like dispersion, approximately below red line the CO can be observed, below blue line the SC takes place) for systems with one (panels (a)-(c)) and two (panels (d),(f)) conducting layers in the unit cell. For all systems with single conducting layer in the unit cell thickness of the (bi)polaron $h$=13.2Å, in (a) $1/\varepsilon^*$=0.25, $R_{bip}^{min} = 7$Å, $m_{\parallel}^{*el}=m_e$; (b) $1/\varepsilon^*$=0.28, $R_{bip}^{min} = 6.52$Å, $m_{\parallel}^{*el}=m_e$; (c) $1/\varepsilon^*$=0.25, $R_{bip}^{min} = 6.31$Å, $m_{\parallel}^{*el}=1.2m_e$; (d) $1/\varepsilon^*$=0.25, $R_{bip}^{min} = 6$Å, $m_{\parallel}^{*el}=m_e$, $h = 6$Å; (f) $1/\varepsilon^*$=0.215, $R_{bip}^{min} = 6.3$Å, $m_{\parallel}^{*el}=1.2m_e, h = 6$Å; inset in (f) $1/\varepsilon^*$=0.16, $R_{bip}^{min} = 8$Å, $m_{\parallel}^{*el}=1.2m_e, h = 4.1$Å. In all the panels $m_{\parallel}^{*h}$=1.2$m_e$, out-of-plane carrier effective mass $m_{\perp}^*$=40$m_e$, distance between two conducting planes in the unit cell (in panels d,f) is 4.1Å, distance between neighbor conducting planes in systems with single conducting plane in the unit cell (panels a-c) is 13.2Å, zero-frequency dielectric constant $\varepsilon_0$=30 (except inset in (f) where $\varepsilon_0$=50), maximum group velocity of phonons $u$=10000 ms$^{-1}$; (e) sketch of experimental phase diagram for YBCO system.

Comparison of Fig.5(a) and (b),(c) shows that at the increase of $1/\varepsilon^*$, or the lattice polarizability, as well as at the increase of the in-plane carrier effective mass the area of existence of the two-liquid system expands both $p$ axis (as in panel (b) and (c) compared with (a)), but the temperature $T_c$ of the SC transition slightly increases only at the increase of $1/\varepsilon^*$. Fig.5(f) illustrates how long the two-liquid system persists upon decreasing the lattice polarizability ($1/\varepsilon^*$ value) at various other parameters of the system. There are also two factors whose (slight) influence is not illustrated by Fig.5. 10 times change of the maximum group velocity of phonons $u$ from $u$=10000 to 1000 ms$^{-1}$, results in approximately 15% decrease in $T^*$ and $T_{CO}$ at fixed $p$, whereas $T_c$ is not affected. Decrease of the (bi)polaron thickness $h$ in a system with two conducting planes in the unit cell (being at the constant distance 4.1Å) from $h$=6Å to 4.1Å results in 10% reduction of all the critical temperatures at given $p$.

Thus, the medium characteristic influencing the superfluid transition temperature is $1/\varepsilon^*$. Increase of $1/\varepsilon^* = 1/\varepsilon_\infty - 1/\varepsilon_0$ may be achieved by increase of $\varepsilon_0$ or decrease of $\varepsilon_\infty$. Typical $\varepsilon_0$ values in cuprates are 30÷50 [44,45], reported value of $\varepsilon_\infty$ in a cuprate is 3.5÷4 [48], the calculated phase diagrams represented in Fig.5 span the interval of $\varepsilon_\infty$ from 3.2 to 5.56. Thus, it seems not easy to achieve essential rise in $T_c$, at least in cuprates, by increasing the lattice polarizability. Likely, more effective way to raise $T_c$ is to affect the spectrum of elementary excitations of the bipolaron liquid by changing the number of conducting layers in the unit cell, interlayer distance, or trying to introduce a gap into the spectrum. As is known spectrum of phason excitations of the CDW can develop a gap in several cases [18]. Initially we calculated $T_c$ using the spectrum of elementary excitations of the large-bipolaron liquid deduced in [43] which has a gap, it yielded presence of superfluid component even at temperature $T=T^*$. In principle, the upper limit of the superconducting transition temperature $T_c$ for a given $p$ in the systems under consideration is the temperature $T^*$, which limits the region of existence of the bipolaron liquid, so that room-temperature superconductivity in strongly interacting electron-phonon systems is, in principle, possible.

Stability of shape of the calculated phase diagram at changing the system parameters demonstrated by Fig.5 is caused by invariance of reasons that control the phase diagram of the two-liquid system. The optimal doping corresponds to reaching the maximum (or almost maximum) bipolaron density for given system parameters as is illustrated by Fig.6(a). Interestingly, that recently similar saturation of energy of 3D plasmons in electron-doped cuprates upon reaching the optimal doping was reported [62]. This may point to possible bipolaron nature of the observed plasmon excitations resembling the amplitudon branch of the CDW elementary excitations [61]. Further increase in $p$ above the optimal doping does not change $T_c$ essentially until the boundary of the region of the bipolaron liquid existence (green line $T^*(p)$) is reached because extra holes occupy delocalized states.

At overdoping, as $p$ increases, sooner or later the free energy of pure Fermi liquid becomes lower than the free energy of a two-liquid system, thus shaping the overdoping side of the phase diagram. The physical reason why high doping makes autolocalized states energetically unfavorable is division of the momentum space between self-trapped and delocalized carriers. When (bi)polaron states are filled the cold hole states are forbidden for delocalized holes. Energy cost of these exclusion is rather high, as is illustrated by Fig.1(b) where the minimal energy of hot hole states is shown with the green lines 3, and it becomes higher with increasing doping due to increase of density of the hot holes (for example, blue line 4 in Fig.1(b) demonstrates FS in presence of bipolarons at $p$=0.25). Besides, the carrier states in near-antinodal regions (areas 2 and 2' in Fig.3(b)) are excluded too.

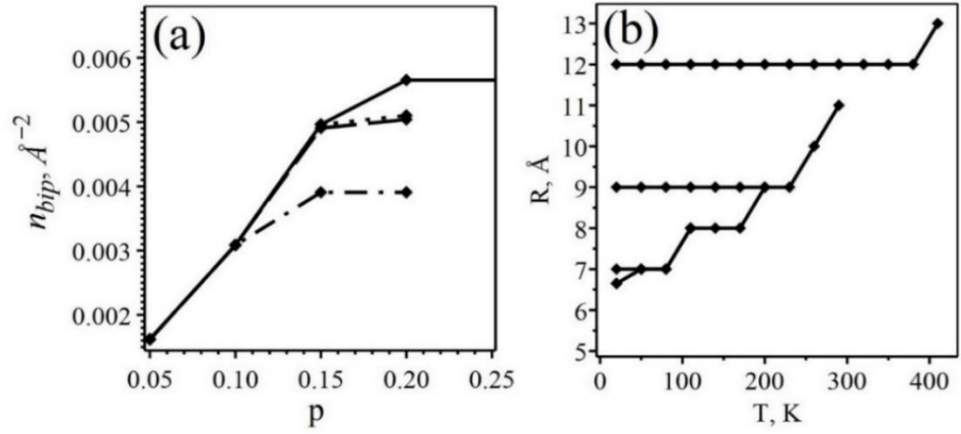

FIG.6. (a) Doping dependence of the calculated bipolaron density at several temperatures (from the top to the bottom $T$=20, 50, 80 and 110K) and (b) calculated equilibrium bipolaron radius as function of temperature (lines from the top to the bottom correspond to $p$=0.05, 0.1 and 0.15 and 0.2) in a system with a single conducting plane in the unit cell and $1/\varepsilon^*$=0.25, $R_{bip}^{min}$ = 6.52Å, $h$=13.2Å, $\varepsilon_0$=30, $m_\parallel^{*el}$=1.2$m_e$, $m_\perp^*$=40$m_e$.

Higher $T^*$ and $T_{CO}$ values at low doping obtained here are due to larger size of (bi)polarons at lower doping as is illustrated by Fig.6(b), larger radius corresponds to higher binding energy (Fig.4(a,b)). Decrease of the bipolaron size at the increase of doping up to the optimal doping results in decrease of $T^*$ and $T_{CO}$ but increases $T_c$ due to increase of density of hole bipolarons. Such behavior is likely realized in La-based systems according to observed in them doping behavior of the CO wave vector $K_{CO}(p)$ shown as green rhombs in Fig.7. However, there is another variant of the system behavior at low doping which is probably realized in Y- and Bi-based cuprates (according to observed in them $K_{CO}(p)$ shown as symbols in Fig.7). It is inhomogeneous system comprising areas of two-liquid system of carriers with bipolaron radius about 6÷8Å and areas of pure Fermi-liquid [34]. In such case increase in $T_c$ with doping results from expansion of area of the conducting plane occupied by two-liquid system of carriers and corresponding rise of the bipolaron density. To obtain such a variant of $K_{CO}(p)$ in the calculation, it is necessary to refine the present approach, as discussed after Fig. 7.

Equilibrium bipolaron size is determined by balance in the free energy. Increasing the bipolaron radius increases the bipolaron binding energy (as is illustrated by Fig.4(a,b)) and decreases the FBZ region excluded due to the CO potential (shown in Fig.3(b)) that lowers the system free energy. But increase of the bipolaron radius decreases the bipolaron density that increases the free energy of the system. Fig.6(b) represents equilibrium bipolaron radius $R_{bip}(T)$ as function of temperature at different doping $p$.

Calculated by minimizing the system free energy equilibrium radius of the bipolaron allows to compute the CO wave vector as [34] $K_{COx} = K_{COy} = 2\pi/(2R_{bip}) = a_0/(2R_{bip})$ (r.l.u.), where $a_0$ is a lattice constant and r.l.u.= $2\pi/a_0$. Fig.7 represents the calculated CO wave vector as function of doping at temperature T=20K for various system parameters. The obtained values of the $K_{CO}$ are in the same range as ones observed in cuprates. Fig.7 shows also interesting feature of the calculated $K_{CO}(p)$ dependences demonstrated by experimental $K_{CO}(p)$ too: at the optimal doping they tend to close values for different systems. In distinct from the phase diagram the shape of the calculated $K_{CO}(p)$ dependence is sensitive to variation of the dispersion and medium parameters, as it occurs in cuprates where systems with different $K_{CO}(p)$ behavior (shown with symbols in Fig.7) demonstrate the same shape of the phase diagram.

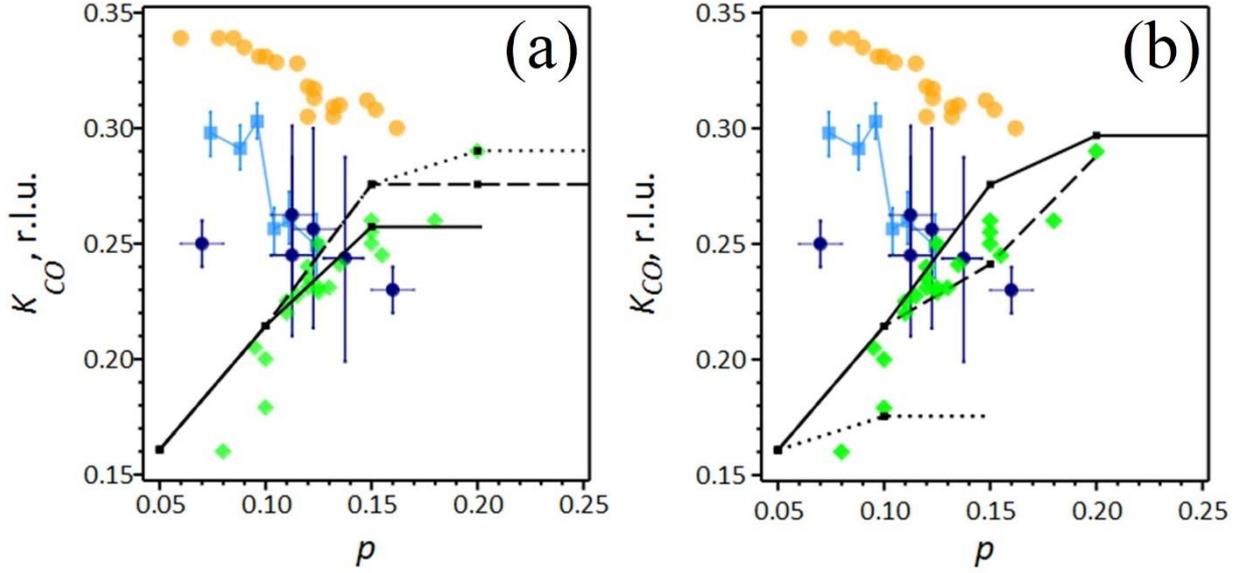

FIG.7. Calculated CO wave vector $K_{CO}$ at the temperature $T=20K$ as function of doping in a systems with $m_\parallel^{*h}=1.2m_e$ $\varepsilon_0=30$ and (a) single conducting plane in the unit cell, $h=13.2$Å, and $1/\varepsilon^*=0.25$, $R_{bip}^{min}=7$Å, $m_\parallel^{*el}=m_e$ (solid line), $1/\varepsilon^*=0.28$, $R_{bip}^{min}=6.52$Å, $m_\parallel^{*el}=m_e$ (dashed line) and $1/\varepsilon^*=0.25$, $R_{bip}^{min}=6.31$Å, $m_\parallel^{*el}=1.2m_e$ (dotted line) and (b) two conducting planes at a distance 4.1Å in the unit cell and $1/\varepsilon^*=0.25$, $m_\parallel^{*el}=m_e$, $R_{bip}^{min}=6$Å, $h=6$Å (solid line); $1/\varepsilon^*=0.215$, $m_\parallel^{*el}=1.2m_e$, $R_{bip}^{min}=6.3$Å, $h=6$Å (dashed line); $1/\varepsilon^*=0.16$, $\varepsilon_0=50$, $m_\parallel^{*el}=1.2m_e$, $R_{bip}^{min}=8$Å, $h=4.1$Å (dotted line). Systems in panels (a) and (b) are the same as in Fig.5 in the first and second rows, respectively. Symbols show some experimental results [3,6,64,65] on YBCO (orange circles), LaSCO (green rhombs), Bi2212 (light blue squares with error bars) and Bi2201 (blue circles with error bars).

At underdoping the calculated $K_{CO}(p)$ dependences are increasing like in LSCO (green rhombs in Fig.7), however other cuprates demonstrate decreasing or almost constant $K_{CO}(p)$ at underdoping. To obtain variety of $K_{CO}(p)$ dependences at underdoping, probably, a refined distribution is needed, in which the density of (bi)polarons is determined not by their binding energy but by difference in the free energy of the system with (bi)polarons and without them, may be a different carrier dispersion is also necessary. The present form of distribution does not yield hole bipolarons with small size (corresponding to large $K_{CO}$) at underdoping since low-size polarons have lower binding energy than bipolarons of the same size, as is seen from Fig.4(a,b). But the total free energy is higher in presence of hole polarons instead of hole bipolarons due to imbalance in the energy of interaction between autolocalized carriers discussed in the next paragraph.

Let us consider why polarons being more profitable energetically than bipolarons at small radius as is demonstrated by Fig.4(a,b) nevertheless are superseded by bipolarons in the doping range considered here. As our analysis shows this occurs due to two reasons. The first reason is the energy of Coulomb repulsion of two polarons with the same coordinate of their centers, this reason hampers existence of electron polarons in hole doped systems where the electron density is high. For the hole polarons the reason is imbalance of energies of interaction between autolocalized carriers with the opposite sign of charge. At low hole doping, a system comprising electron bipolarons and hole polarons, it would seem, should be energetically more profitable than bipolarons of both sign of charge. However, in the former system increase of the free energy due to uncompensated repulsion of electron bipolarons (whose charge is twice the polaron charge) outweighs its decrease due to higher binding energy of polarons.

It is interesting to note that besides known $k_F$ mismatch (particle-hole asymmetry) peculiarity of the PG in cuprates [4,5] which is also present in the calculated ARPES spectra of two-liquid system of charge carriers [42] there is another particle-hole asymmetry in such systems at doping lower than optimal. Below the optimal doping in the considered system the hole states near the FS are autolocalized whereas electron states near the FS are delocalized. This asymmetry may display itself in STM experiments. For example, CO will be observed in tunneling experiment only for tunneling of electrons into the sample as it was observed in a cuprate [6]. At doping higher than the optimal one, hole states near the FS are delocalized too so that CO may cease to be observable in tunneling experiments even for tunneling of electrons into the sample like it occurs in experiment [6]. X-ray scattering (RXS) experiments observe CO not only near the FS so that it can observe CO at doping higher than the optimal [8,9]. It is also important to note that decrease in the intensity of the CO peak observed with REXS method below $T_c$ is only 15% (from 1.15 to 1 of normalized units) [6] so that CO in cuprates coexists with the SC. Besides, fluctuations of phase of the order parameter below the temperature of the superfluid transition in the bipolaron liquid being the fluctuations of phase of the CDW may hamper observation of the CO in the regions of presence of the superfluid component [59].

The calculated density $n_s$ of the superfluid component in the overdoping region decreases down to zero with increasing doping, as is illustrated by Fig.7, in consent with the experiments on cuprates [58,63]. These experiments highlight the distinct of cuprates from the BCS systems where increase of the carrier density results in increase of the phase stiffness and $T_c$. The slope of the calculated $n_s(p,T)$ on the overdoping side of the phase diagram may be not so steep as in Fig.8, if a more refined method of the bipolaron density $n_{bip}$ calculation will be developed. It should describe a gradual decrease of $n_{bip}$ at approaching the temperature at which free energies of two phases - the phase with (bi)polaron liquid and delocalized carriers and the phase with delocalized carriers only - coincide (at given doping). In the present calculation the bipolaron density is determined by the bipolaron binding energy (together with the doping level) and sharply falls to zero when the energy of system with bipolarons becomes higher than the energy of the system without them (if this occurs at lower temperatures than the thermal destruction of bipolarons at given doping).

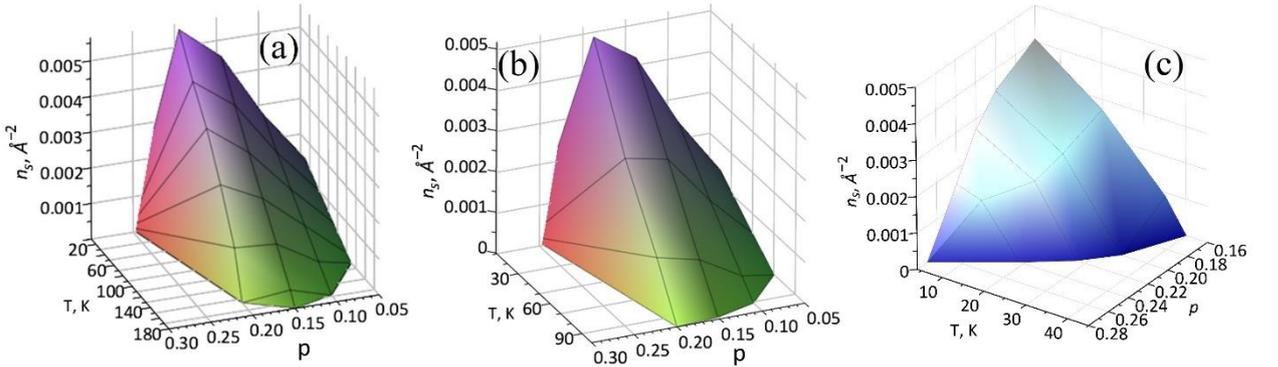

FIG.8. Calculated superfluid density $n_s(p,T)$ as function of doping and temperature (a) in a system with two conducting planes in the unit cell and $R_{bip}^{min} = 6$Å, $h$=6Å, $m_\parallel^{*el}$=$m_e$, and (b) in a system with one conducting plane in the unit cell and $R_{bip}^{min} = 6.31$Å, $h$=13.2Å, $m_\parallel^{*el}$=$1.2m_e$, in both (a) and (b) $1/\varepsilon^*$=0.25, $\varepsilon_0$=30, $m_\parallel^{*h}$=$1.2m_e$, $m_\perp^*$=$40m_e$; (c) doping and temperature dependence of density of the superconducting condensate computed from the experimentally measured phase stiffness in LaSrCuO system [58].

As was mentioned above the mass of the supercurrent carriers does not coincide with the effective mass of individual bipolarons. As supercurrent in the considered system is related with a propagation of the polarization wave drugging the autolocalized carriers the effective mass of the supercurrent carriers is much lower than the effective mass of individual bipolarons.

## VI. CONCLUSION AND OUTLOOK

In summary, here we consider finite-temperature states of strongly coupled carrier (with cuprate-like dispersion) and phonon fields and conditions of existence of two-liquid system of charge carriers. We found that in the phase diagram doping-temperature of strongly coupled electron and phonon fields there is a region where two-liquid system of charge carriers comprising large-bipolaron liquid and Fermi liquid exists. As a consequence, the areas of the phase diagram with presence of superconductivity (superfluidity of the bipolaron liquid), charge order (local order in the bipolaron liquid) and pseudogap (a consequence of changing the stationary states of delocalized carriers in presence of the bipolarons potential and hole-like dispersion) are identified. Their places in the phase diagram coincide with those observed in cuprates. Thus, we show that there is at least one class of systems in which PG, CDW and SC have a common nature, as was recently assumed for cuprates [7].

The suggested approach has allowed identifying physical reasons shaping both sides of the phase diagram (underdoped and overdoped) and determining the optimal doping (where the bipolaron density reaches the maximum value). Increase of the plasmon energy with doping and its saturation at the optimal doping [62] and decrease of the superfluid density with doping at overdoping [58] naturally appear in this approach. Predicted in the approach relations between the CO wave vector and the pair size, or equilibrium bipolaron size, the pair size and the optimal doping, the CO wave vector and the wave vector of the waterfall in ARPES spectrum are observed in cuprates. The range of the calculated CO wave vector values coincides with occurring in cuprates (at using characteristic of them values of the static and high-frequency dielectric constant [44,45,48] and carrier effective mass).

Practical significance of the results obtained is to chart ways of controlling the properties of systems with bipolaron condensate, with the hope of achieving the same high level of control as was realized for excitonic condensates [66,67]. We reveal the system characteristics influencing the phase diagram doping - temperature of strongly coupled carrier (with cuprate-like dispersion) and phonon fields and, in particular, the temperature of the superconducting transition. In principle, the upper limit of the superconducting transition temperature $T_c$ in the considered systems is the temperature $T^*$ bounding the area of the bipolaron liquid existence, so that room temperature superconductivity in strongly interacting electron-phonon systems is possible. At given medium parameters, the effective way to operate $T_c$ is by changing the spectrum of elementary excitations of the large-bipolaron liquid for example by trying to introduce a gap into it like one that may occur in the spectrum of CDW elementary excitations.

From the theoretical point of view the results add a missing element to the condensed matter physics by revealing the finite-temperature states of systems with strong long-range (Fröhlich) EPI and high carrier density (earlier only tasks about strong short-range EPI at high carrier density were solved [31,68,69]) and the existence conditions for the two-liquid system of charge carriers and superfluid component of the large-bipolaron liquid. This opens possibilities for further development of the field. For example, a study of coherent motion of a condensate of coupled carrier and polarization field promises many interesting features, which are the subject of a separate article. Obviously, the current emerges at the condensate motion since the electrons and holes belong to the same band as was noted earlier for the CDW case [17]. The symmetry of the phase of the condensate is probably related with the symmetry of the lattice deformation in the polarization (charge density) wave which is d-wave for the cuprates [12] but may be different in other systems. Coherent propagation of the polarization wave during the motion of the condensate of the bipolar liquid significantly reduces the effective mass of supercurrent carriers in comparison with the effective mass of individual bipolarons.

**Acknowledgements**. The authors are grateful to E. I. Shneider for useful comments and to S. B. Rochal for many discussions.